\documentclass[conference,hidelinks]{IEEEtran}
\IEEEoverridecommandlockouts
\usepackage{cite}
\usepackage{amsmath,amssymb,amsfonts}
\usepackage{algorithmic}
\usepackage{textcomp}
\usepackage{xcolor}
\usepackage{graphicx}
\usepackage{multirow}
\usepackage{makecell}
\usepackage{csquotes}
\usepackage{framed}
\usepackage{array}
\usepackage{dblfloatfix}
\usepackage[bookmarks=false]{hyperref}
\usepackage[capitalise]{cleveref}

\newcolumntype{L}[1]{>{\raggedright\let\newline\\\arraybackslash\hspace{0pt}}m{#1}}
\newcolumntype{C}[1]{>{\centering\let\newline\\\arraybackslash\hspace{0pt}}m{#1}}
\newcolumntype{R}[1]{>{\raggedleft\let\newline\\\arraybackslash\hspace{0pt}}m{#1}}

\makeatletter
\renewcommand{\@IEEEsectpunct}{\ \,}%
\makeatother

\def\BibTeX{{\rm B\kern-.05em{\sc i\kern-.025em b}\kern-.08em
    T\kern-.1667em\lower.7ex\hbox{E}\kern-.125emX}}

\begin{document}

\title{\huge Automatically Identifying Relations Between Self-Admitted Technical Debt Across Different Sources
}

\author{Yikun Li, Mohamed Soliman, Paris Avgeriou \\
\IEEEauthorblockA{Bernoulli Institute for Mathematics, Computer Science and Artificial Intelligence \\
University of Groningen \\
Groningen, The Netherlands \\
\{yikun.li, m.a.m.soliman, p.avgeriou\}@rug.nl}
}

\maketitle

\begin{abstract}
\emph{Self-Admitted Technical Debt} or \emph{SATD} can be found in various sources, such as source code comments, commit messages, issue tracking systems, and pull requests.
Previous research has established the existence of relations between SATD items in different sources; such relations can be useful for investigating and improving SATD management.
However, there is currently a lack of approaches for automatically detecting these SATD relations.
To address this, we proposed and evaluated approaches for automatically identifying SATD relations across different sources.
Our findings show that our approach outperforms baseline approaches by a large margin, achieving an average F1-score of 0.829 in identifying relations between SATD items. 
Moreover, we explored the characteristics of SATD relations in 103 open-source projects and describe nine major cases in which related SATD is documented in a second source, and give a quantitative overview of 26 kinds of relations.
\end{abstract}


\section{Introduction}

The implicit documentation of technical debt by developers with statements like ``TODO'' or ``FIXME'' has become known as \emph{Self-Admitted Technical Debt} or \emph{SATD} \cite{potdar2014exploratory}.
This can take place in various sources such as source code comments, commit messages, issue tracking systems, and pull requests \cite{zampetti2021self}.
The analysis of SATD has shown to be an important and valuable complement to using static source code analysis for technical debt management \cite{SIERRA201970}.
Consequently, significant research work has focused on identifying SATD from various sources; most of this work uses source code comments \cite{da2017using,ren2019neural}, but recently some work also used issue tracking systems \cite{li2022identifying}, commit messages \cite{zampetti2018self}, and pull requests \cite{li2022automatic}.

Furthermore, recent research has found that there are relations between SATD items in different sources\cite{zampetti2018self,li2022automatic}.
For example, developers may document and explain the same SATD item in both code comments and issues, and then record its repayment in a commit message. The automated identification of such SATD relations could support more efficient and effective SATD management. First, it would allow developers to trace multiple SATD items at once.
For instance, tracing the SATD from code comments to issues could assist developers in identifying other relevant SATD items under the same issue.
As a result, developers could address these related SATD items together.
Furthermore, it could assist developers in having a better understanding of how debt is introduced, accumulated, and eventually repaid.
This would give a better overview of how SATD evolves in the system. Last, but not least, it  would help to tackle individual SATD items by combining information from multiple sources and forming a more complete picture on how to prioritize and fix them. 

Despite this significance of SATD relations in different sources,  there is currently a lack of approaches for automatically identifying such relations.
To address this problem, we propose and evaluate approaches for automatically identifying relations between instances of SATD across four different sources: source code comments, issue trackers, commit messages, and pull requests.
Specifically, as related SATD items are often explicitly linked (e.g. a commit message references a pull request through its unique ID - see \cref{f:links}), we use such linked data between different sources and randomly collect a dataset of 1,000 pairs of SATD items from 103 open-source projects. 
We then manually analyze these pairs to identify the relations between SATD items.
Using this dataset for training, we compare the predictive performance of two deep learning approaches (i.e., BERT-based and CNN-based) and one baseline method (i.e., random) for automatically identifying SATD relations. 
Finally, we summarize the characteristics of SATD relations with examples and present the number of various types of SATD relations in 103 open-source projects.
The main contributions of this paper are as follows:

\begin{itemize}
    \item \textbf{Contributing a self-admitted technical debt relation dataset.}
    We gather a dataset of 1,000 pairs of SATD items from 103 open-source projects across four sources (code comments, commit messages, pull requests, and issues).
    Our dataset includes manual annotations on the relationship types of each SATD pair.
    In order to promote further research in this area, we make our dataset publicly accessible\footnote{\label{l:data}\url{https://github.com/yikun-li/satd-relations}}.

    \item \textbf{Proposing an approach to automatically identify relations between SATD items.}
    We propose a BERT-based approach for identifying SATD relations and compare it to a CNN-based approach and a baseline method in terms of F1-score.
    The results showed that our BERT-based approach attains an average F1-score of 0.829 for identifying two types of SATD relations, significantly outperforming the CNN-based approach and the baseline.

    \item \textbf{Exploring the amount of data necessary for training the model.}
    We find that only a small amount of training data is needed to achieve a high level of accuracy in identifying relations between SATD items.

    \item \textbf{Characterizing SATD relations in 103 open-source projects.}
    We report on the sequences and amount of SATD relations.
    Specifically, with the aim of gaining a \textit{qualitative} understanding of SATD relations, we summarize the most prevalent SATD relations and provide examples based on the manually annotated dataset.
    To provide a better overview of the \textit{quantity} of SATD relations, we then apply the proposed machine learning model to automatically identify all relations between SATD items from 103 open-source projects.
    We consequently present the number of various types of SATD relations.
\end{itemize}

The structure of the remainder of this paper is as follows.
In \cref{sec:related}, we discuss related work. 
The case study design is then described in detail in \cref{sec:approach}, and the results are presented and discussed in \cref{sec:results} and \cref{sec:discussion}, respectively.
In \cref{sec:validity}, we evaluate threats to validity, and we draw conclusions in \cref{sec:conclusion}.

\section{Related Work}
\label{sec:related}

In this work, we focus on automatically identifying relations between SATD items across different sources.
Thus, we review related work from two areas: 1) research on SATD in general and 2) research on SATD relations.

\subsection{Self-Admitted Technical Debt.}

Potdar and Shihab \cite{potdar2014exploratory} were pioneers in the field of studying SATD in source code comments.
They conducted an analysis of four open-source projects, identifying instances of SATD within the codebase.
Their findings revealed that between 2.4\% and 31\% of source files contained SATD comments, and that only between 26.3\% and 63.5\% of identified SATD was removed after its introduction.
Then Maldonado and Shihab \cite{maldonado2015detecting} expanded on this work by focusing on the different types of SATD present in open-source projects.
They analyzed 33,000 code comments from five projects, categorizing SATD into five distinct types: design, requirement, defect, documentation, and test debt.
The study results indicated that design debt was the most prevalent type of SATD, accounting for between 42\% and 84\% of the classified instances.

Following the initial work on exploring SATD in source code comments, there has been a significant focus on developing automatic methods for SATD identification. 
Maldonado \textit{et al.} \cite{da2017using} manually classified source code comments from ten open-source projects into different types of SATD and used a maximum entropy classifier to automatically identify design and requirement debt.
Moreover, Ren \textit{et al.} \cite{ren2019neural} proposed a convolutional neural network-based approach to accurately identify SATD from source code comments.
Some researchers have also focused on identifying SATD from other sources.
Li \textit{et al.} \cite{li2022identifying} examined 23,000 issue sections and proposed a convolutional neural network-based approach for identifying SATD from issue tracking systems.
Furthermore, they conducted another study to identify four types of SATD from multiple sources, including source code comments, commit messages, pull requests, and issues \cite{li2022automatic}.

In addition to identifying SATD, there has been research about the measurement, prioritization, and repayment of SATD.
Kamei \textit{et al.} \cite{kamei2016using} examined ways to measure the interest accrual of SATD and proposed to use lines of code (LOC) and Fan-In measures.
Mensah \textit{et al.} \cite{mensah2018value} developed an SATD prioritization scheme that consists of identification, examination, and estimation of rework effort.

\subsection{Relation Between SATD Items.}

There are limited studies that have explored or utilized the relations between SATD items across different sources.
Zampetti \textit{et al.} \cite{zampetti2018self} investigated the degree to which the removal of SATD in code comments is documented in commit messages.
They found that the removal of SATD was documented in commit messages in 131 out of 997 cases.
Li \textit{et al.} \cite{li2022self,li2022automatic} further examined the relations between SATD items across different sources.
Their results indicated that there are four types of relations between SATD in different sources and that the number of related SATD items in different sources is comparable.

Compared to the previously mentioned studies, our study is the first to propose an automated method for identifying relations between SATD items across different sources.
This can help to improve the effective management of technical debt.
Additionally, we applied our trained machine learning model to 103 open-source projects and summarized the characteristics of SATD relations, allowing for a comprehensive understanding of SATD relations and providing valuable insights for developers and managers.

\section{Study Design}
\label{sec:approach}

\begin{figure*}[thpb]
  \centering
  \includegraphics[width=\linewidth]{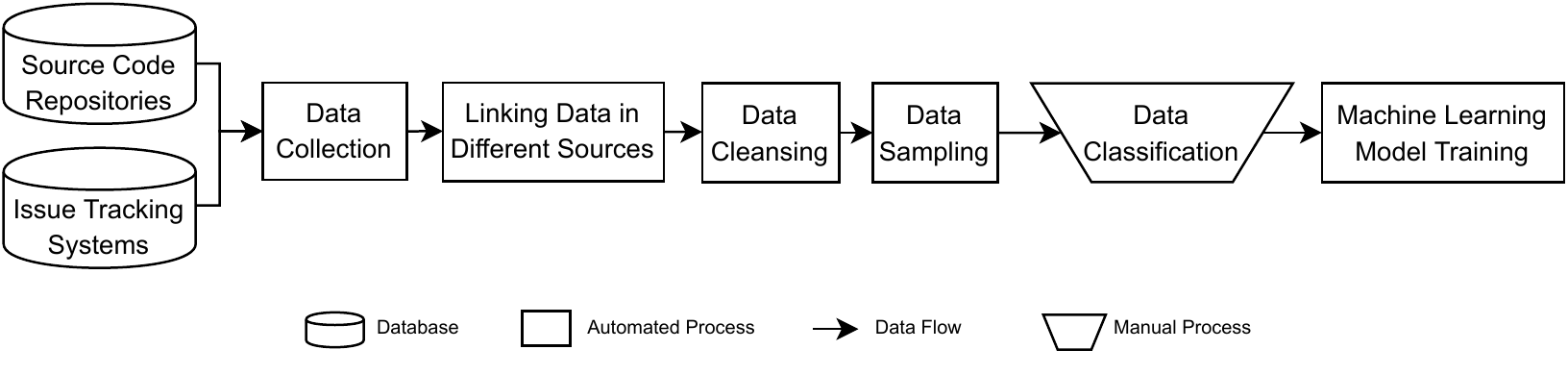}
  \caption{The framework of our study.}
  \label{f:framework}
\end{figure*}

The goal of this study, formulated according to the Goal-Question-Metric \cite{Solingen:02} template is to ``\textit{\textbf{analyze} self-admitted technical debt from source code comments, commit messages, pull requests, and issue trackers \textbf{for the purpose of} automatically identifying and characterizing relations between self-admitted technical debt \textbf{with respect to} the identification accuracy, the sequences of relations, and the quantity of relations \textbf{from the point of view of} software engineers \textbf{in the context of} open-source software systems.}''
This goal is refined into two research questions (RQs):

\begin{itemize}
    \item \textbf{RQ1:} \textit{How to automatically and accurately identify relations between SATD from different sources?}\\
    This RQ is further refined into three sub-RQs:

    \begin{itemize}
        \item \textbf{RQ1.1:} \textit{Which deep learning approaches can effectively identify relations between SATD items?}\\
        \textbf{Rationale:}
        As there is no approach available for SATD relation identification, we investigate the efficacy of two deep learning methods: a CNN-based approach and a BERT-based approach.
        We choose these two methods because: a) previous studies \cite{ren2019neural, li2022automatic} have demonstrated that CNN-based approaches are capable of extracting features from SATD items; and b) the self-attention mechanism of BERT, is particularly powerful in extracting relationships between items \cite{devlin2018bert}.

        \item \textbf{RQ1.2:} \textit{How to further improve the predictive performance of the machine learning model?}\\
        \textbf{Rationale:}
        In order to make machine learning models as accurate and reliable as possible for real-world applications, we need to explore further optimizing the predictive performance of machine learning models.
        This can be achieved, for example, by investigating the effectiveness of pre-training models on domain-specific data (e.g., pre-training on SATD comments).
        
        \item \textbf{RQ1.3:} \textit{How much data is sufficient for training the machine learning models to identify related SATD?}\\
        \textbf{Rationale:}
        The lack of training data could lead to poor machine-learning model performance.
        Therefore, we need to investigate whether the current training dataset is sufficient for training machine learning models to identify SATD relations or whether it needs to be further extended.
    \end{itemize}
    
    \item \textbf{RQ2:} \textit{What are the sequences and quantities of SATD relations?}
    This RQ is further refined into two sub-RQs:
    
    \begin{itemize}
        \item \textbf{RQ2.1:} \textit{What are the sequences of documenting related SATD in different sources?}\\
        \textbf{Rationale:}
        SATD items can be related between two sources in a bi-directional way: developers can consider SATD that already exists in source A and subsequently document it in source B, and vice versa.
        For instance, existing SATD from issues is sometimes also documented in code comments, so that developers are aware of its location; but also, code comments that contain SATD are sometimes also documented as issues, so that they become visible at the issue tracker level.
        Thus, it is important to know the sequences of documenting related SATD.
        This could aid in further understanding how related SATD occurs and is utilized.
        
        \item \textbf{RQ2.2:} \textit{How much SATD is related between different sources?}\\
        \textbf{Rationale:}
        Quantifying how much SATD is related across different sources can help us to understand how developers make use of different sources to manage SATD.
        For example, when developers create an issue to solve a technical debt item, they could document the repayment of that item in commit messages or in code comments.
        Knowing the amount of SATD repayment documented in commit messages or in code comments helps us understand how developers are managing SATD.
        This RQ also gives us insights on how to use SATD relation information.
        For example, if we find that a large amount of SATD in issues, is additionally documented in code comments, we could combine the information in both sources to obtain a more complete understanding of such SATD items.
    \end{itemize} 
\end{itemize}

To answer these Research Questions, we follow the approach that is presented in \cref{f:framework}.
In the following subsections, we will go into more detail about each step.

\subsection{Data Collection}
\label{sec:DataCollection}
To analyze relations between SATD in different sources, we chose to use the dataset provided by the work of Li \textit{et al.} \cite{li2022automatic}.
The reasons for using this dataset are listed below:

\begin{itemize}
    \item This dataset collects data from 103 projects from the Apache ecosystem.
    These projects possess a high level of quality and are maintained by mature communities.
    
    \item This dataset contains 23.6M code comments, 1.2M commit messages, 3.1M issue sections, and 2.5M pull request sections; these provide us with sufficient data to mine SATD items and study SATD relations from different sources.
    We note that because issues and pull requests are both composed of three sections (i.e. summaries, descriptions, and comments), we consider each of these sections individually. This is because developers commonly share a single item of SATD in an issue or pull request section \cite{li2022automatic}.
    In other words, a pull request or issue may have one, two, or more SATD items in the respective sections.
\end{itemize}

The details of this dataset are presented in \cref{tb:projects}.

\begin{table}[htb]
\caption{The minimum, maximum, mean, median, and sum of the number of SATD items for all projects in the dataset.}
\label{tb:projects}
\begin{center}
\resizebox{\columnwidth}{!}{
\def\arraystretch{1.35}
\begin{tabular}{cccccc}
\hline
Source & Min & Max & Mean & Median & Sum \\
\hline
Comments & 326 & 3,894,056 & 229,705 & 80,211 & 23,659,650 \\
Commits & 573 & 70,861 & 12,120 & 6,477 & 1,248,324 \\
Pull Sections & 239 & 636,518 & 24,542 & 8,374 & 2,527,803 \\
Issue Sections & 856 & 303,608 & 21,144 & 5,010 & 3,065,815 \\
\hline
\end{tabular}
}
\end{center}
\end{table}

\subsection{Linking Data in Different Sources}

\begin{figure}[htb]
  \centering
  \includegraphics[width=\linewidth]{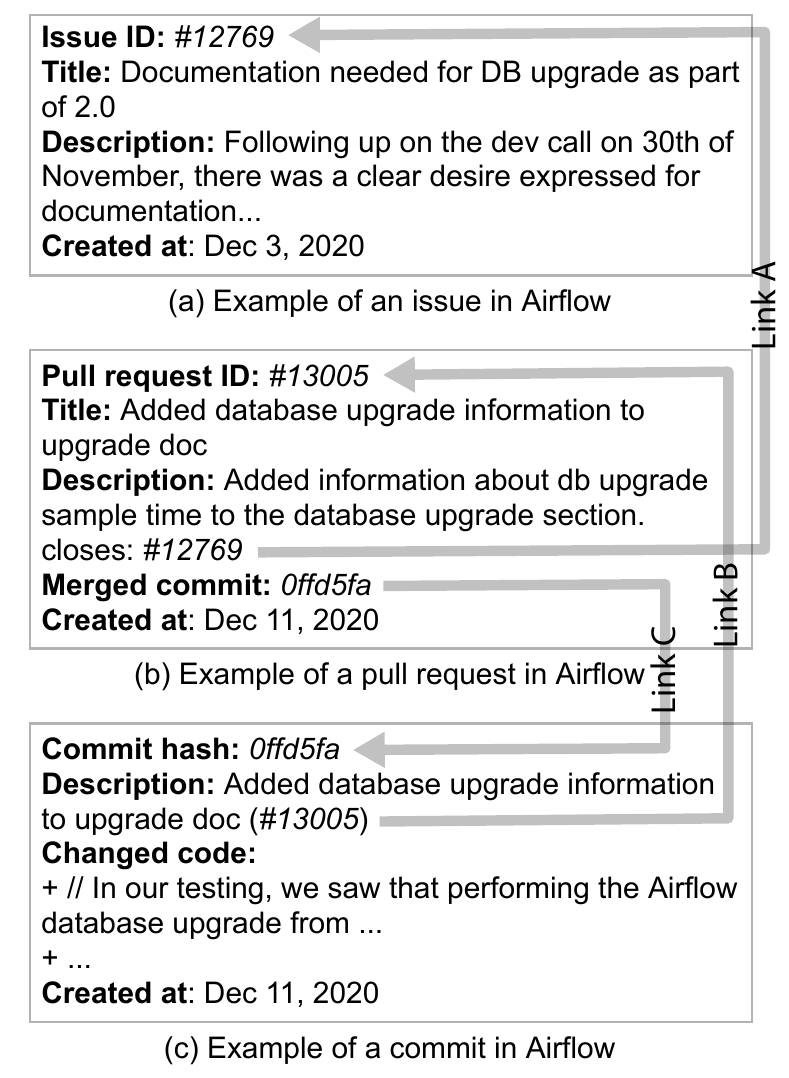}
  \caption{Example links between three sources in Airflow (adapted from \cite{li2022automatic}).}
  \label{f:links}
  \vspace{2mm}
\end{figure}

To study relations between SATD in different sources, we linked different sources using issue ID, pull request ID, and commit hash. This approach is similar to the study of Li et al. \cite{li2022automatic}.
An example of link building is shown in \cref{f:links}.
In this example, we can link the pull request and issue with the issue ID (\emph{\#12769}) (see \emph{Link A}).
Moreover, the pull request and the commit are linked with the pull request ID (\emph{\#13005}) (see \emph{Link B}).
Furthermore, we can also use the commit hash (\emph{0ffd5fa}) to link the commit and pull request (see \emph{Link C}).

\subsection{Data Cleansing}

Some comments are generated by bot users.
Because we focused on technical debt that was explicitly documented and admitted by developers (i.e., SATD), we removed data generated by bot users when manually analyzing the data.
Specifically, during data analysis, when we found that a comment is generated by a bot, we note down the username of the bot.
Then we remove all the comments created by identified bot users.

\subsection{Data Sampling}

Li \textit{et al.} observed in their study \cite{li2022automatic}, that whether two SATD items are related, is associated with the cosine similarity between the two items.
Specifically, a higher cosine similarity indicates a higher likelihood of SATD items being related, with most SATD items considered related when the cosine similarity is greater than 0.5.
Moreover, they also discovered that the majority of the SATD item pairs have low cosine similarity, with an average similarity score of 0.135 and a standard deviation of 0.142.
If we randomly select pairs of SATD items from the dataset, it is likely that most of the pairs will have low similarity, thus resulting in a very small percentage of SATD items being related.
To solve this issue, we used the stratified sampling method, similar to the previous study \cite{li2022automatic}, to obtain 10 groups of samples with similarity scores between 0 and 1: 0.1 and 0.2, ..., 0.9 and 1.
This way of sampling ensures the diversity of the sample for analysis.
In total, we obtained a sample of 1,000 pairs of SATD items to be analyzed.
It is worth noting that in RQ1.3, we investigate whether this amount of data is indeed sufficient to train the machine learning models.

\subsection{Data Classification}
\label{sec:DataClassification}

In order to construct a dataset of SATD relations for training machine learning models, we first developed a classification framework through the independent analysis of 50 pairs of SATD items by the first and second authors.
The results of this analysis were then compared and discussed between the two authors to ensure consistency and agreement on the classification framework.
To further refine the framework, the first and second authors independently classified an additional 50 pairs of SATD items, resulting in a more consistent, accurate, and condensed classification framework.
Finally, we reached an agreement on the framework, comprising two types of relationships: \textbf{duplication of SATD} referring to documenting identical SATD items in different sources; and \textbf{repayment of SATD}, referring to documenting in source A the repayment of an SATD item, originally reported in source B.
For example, after documenting an SATD item in code comments, developers may occasionally report it in issues.
We categorize this case as \emph{duplication of SATD} (see more examples in \cref{sec:2.1}).
The first author then independently classified 1,000 pairs of SATD items.
Of these pairs, 197 were classified as SATD duplication and 390 pairs were classified as SATD repayment.
To further assess the level of agreement on the annotations, a random sample of 15\% of the dataset (150 pairs) was selected and independently classified by the second author.
Cohen's kappa coefficient \cite{landis1977measurement}, which measures the level of agreement between the classifications of the first and second authors, was calculated.
The results indicate that we achieved a \emph{substantial} level of agreement \cite{landis1977measurement} with a coefficient of $+0.785$.
Examples of SATD relations are provided in \cref{sec:2.1}.

\begin{figure}[t]
  \centering
  \includegraphics[width=\linewidth]{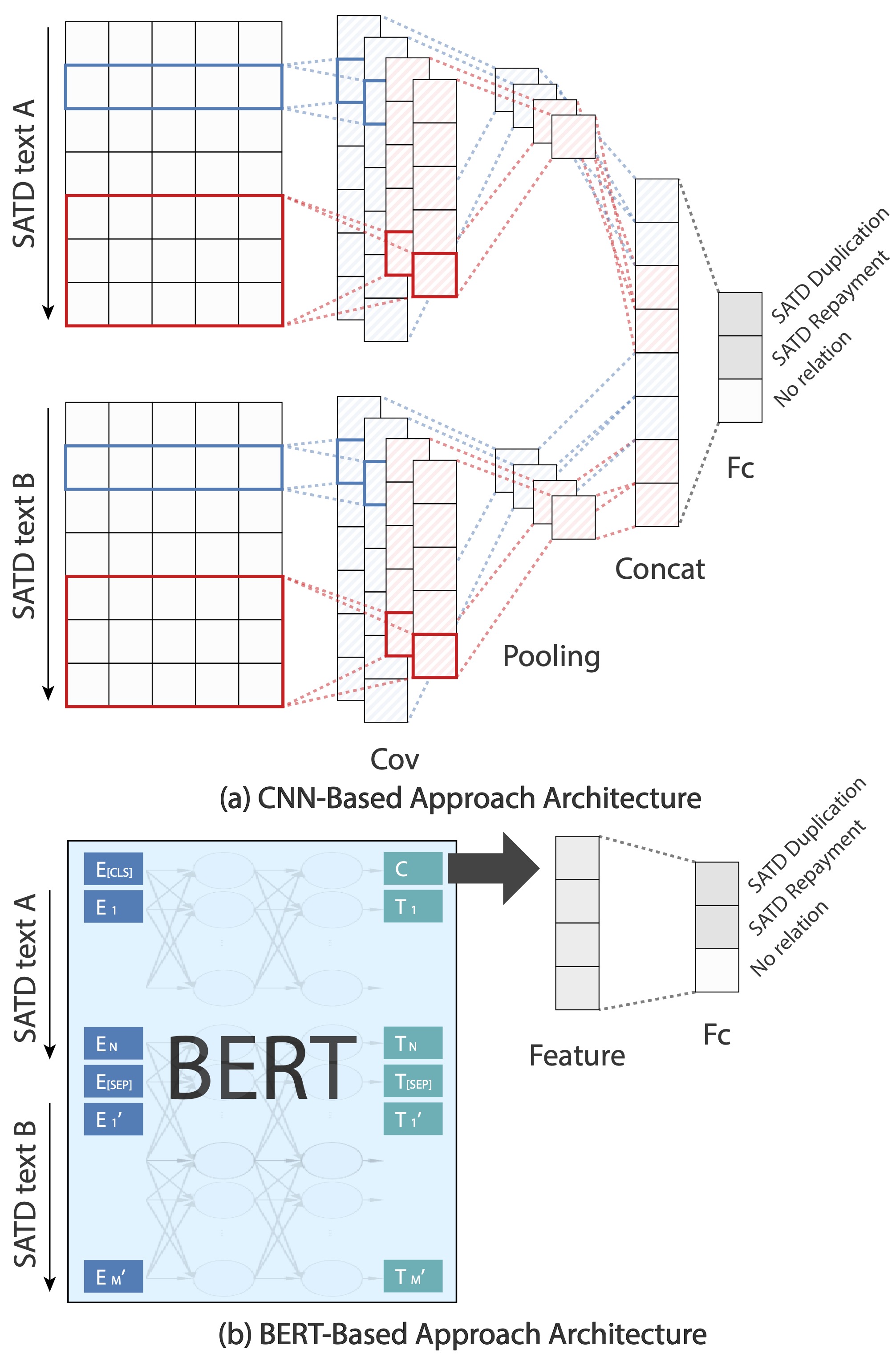}
  \caption{Architecture of machine learning models.}
  \label{f:architecture}
  \vspace{2mm}
\end{figure}

\subsection{Machine Learning and Model Training}

We propose two deep-learning approaches for automatically identifying relations between pairs of SATD items: BERT-based and CNN-based.
In order to provide an objective evaluation of the effectiveness of our proposed approaches for SATD relation identification, we compare their performance with one random baseline approach.
The details of these three approaches are presented below:

\begin{itemize}
    \item \textbf{CNN-based Approach}:
    Text-CNN is a state-of-the-art text classification algorithm proposed by Kim \cite{kim2014convolutional} that has been widely used in several studies on SATD identification \cite{ren2019neural,li2022identifying}.
    Given its demonstrated effectiveness in identifying SATD, we leveraged Text-CNN to extract features for SATD relation identification in this study. 
    Inspired by the Siamese network architecture, we created a Siamese Text-CNN architecture to identify relations between SATD items.
    The architecture of Text-CNN is demonstrated in \cref{f:architecture} (a).
    As can be seen, our CNN-based approach utilizes Text-CNN to extract features from two input SATD items, concatenates the extracted features, and then predicts the type of relationship between the two SATD items.
    
    \item \textbf{BERT-based Approach}:
    BERT is a pre-trained transformer-based neural network that has been widely utilized for natural language processing tasks such as text classification, question answering, and language translation \cite{devlin2018bert}.
    BERT has achieved state-of-the-art performance on various natural language understanding tasks by learning deep bidirectional representations of text.
    Given that BERT leverages the self-attention mechanism, which helps the model to understand the relationships between different words in a sentence and how they depend on each other, it is particularly adept at capturing underlying relations in natural language.
    Therefore, we employed the BERT base model to extract features for identifying relations between SATD items.
    As we can see in \cref{f:architecture} (b), a fully-connected layer is added after feature extraction with the BERT model to predict the SATD relation types.
    
    \item \textbf{Random Approach}:
    This baseline approach randomly predicts the SATD relations based on the probability of the proportion of SATD relation types present in our SATD relation dataset.
\end{itemize}

In this study, we implemented machine learning approaches utilizing the Pytorch library.
For the CNN-based approach, we set the word-embedding dimension at 300, employed filters of five different window sizes (1, 2, 3, 4, 5), and utilized 200 filters for each window size.
This configuration was previously found to achieve the best F1-score in identifying SATD in previous research \cite{li2022automatic}.
Furthermore, we utilized the pre-trained BERT base model (uncased) for feature extraction.
All machine learning models were trained on NVIDIA Tesla V100 GPUs.

\section{Results}
\label{sec:results}

\subsection{RQ1: How to automatically and accurately identify relations between SATD from different sources?}

\subsubsection{\textbf{RQ1.1: Which deep learning approaches can effectively identify relations between SATD items?}}
\label{sec:1.1}

To accurately evaluate the predictive performance of machine learning approaches, we choose to use stratified 10-fold cross-validation.
Specifically, we shuffle and split the manually-analyzed dataset into ten equally-sized partitions, while keeping the number of duplicated SATD pairs and repayment SATD pairs (see Section \ref{sec:DataClassification}) in each division approximately equal.
Then, we choose one of the ten subgroups for testing and the remaining nine for training, repeating the process for each subset and calculating the average precision, recall, and F1-score across ten experiments.

\begin{table}[thb]
\caption{Comparison of the predictive performance of the 3 approaches. Results are averaged over 3 random initializations.}
\label{tb:training_results}
\begin{center}
\resizebox{\columnwidth}{!}{
\def\arraystretch{1.35}
\begin{tabular}{C{1.5cm}C{2.2cm}C{0.7cm}C{0.7cm}C{1.0cm}}
\hline
\textbf{Approach} & Type of Relation & Pre. & Rec. & F1-score \\
\hline
\multirow{3}{*}{\makecell[c]{\textbf{BERT\textsubscript{BASE}} \\(as features)}} & SATD Duplication & 0.786 & 0.743 & 0.759 \\
& SATD Repayment & 0.839 & 0.870 & 0.853 \\
\cline{2-5}
& Average & \textbf{0.813} & \textbf{0.807} & \textbf{0.806} \\
\hline
\multirow{3}{*}{\textbf{CNN}} & SATD Duplication & 0.610 & 0.453 & 0.508 \\
& SATD Repayment & 0.739 & 0.684 & 0.706 \\
\cline{2-5}
& Average & 0.674 & 0.568 & 0.607 \\
\hline
\multirow{3}{*}{\makecell[c]{\textbf{Random} \\(baseline)}} & SATD Duplication & 0.147 & 0.446 & 0.196 \\
& SATD Repayment & 0.310 & 0.307 & 0.227 \\
\cline{2-5}
& Average & \underline{0.235} & \underline{0.408} & \underline{0.234} \\
\hline
\end{tabular}
}
\end{center}
\end{table}

\cref{tb:training_results} shows the precision, recall, and F1-score of the two deep learning approaches (BERT-based and CNN-based) and the baseline approach (random).
As can be seen, the BERT-based approach achieved the highest F1-score of 0.759 and 0.853, for identifying SATD duplication and SATD repayment respectively, with an average F1-score of 0.806.
Moreover, we notice that the CNN-based method achieved a lower average F1-score of 0.607, compared to the BERT-based approach.
Both of these deep-learning approaches significantly outperform the average F1-score of the random baseline (0.234).

\begin{figure*}[t]
  \centering  
  \includegraphics[width=\linewidth]{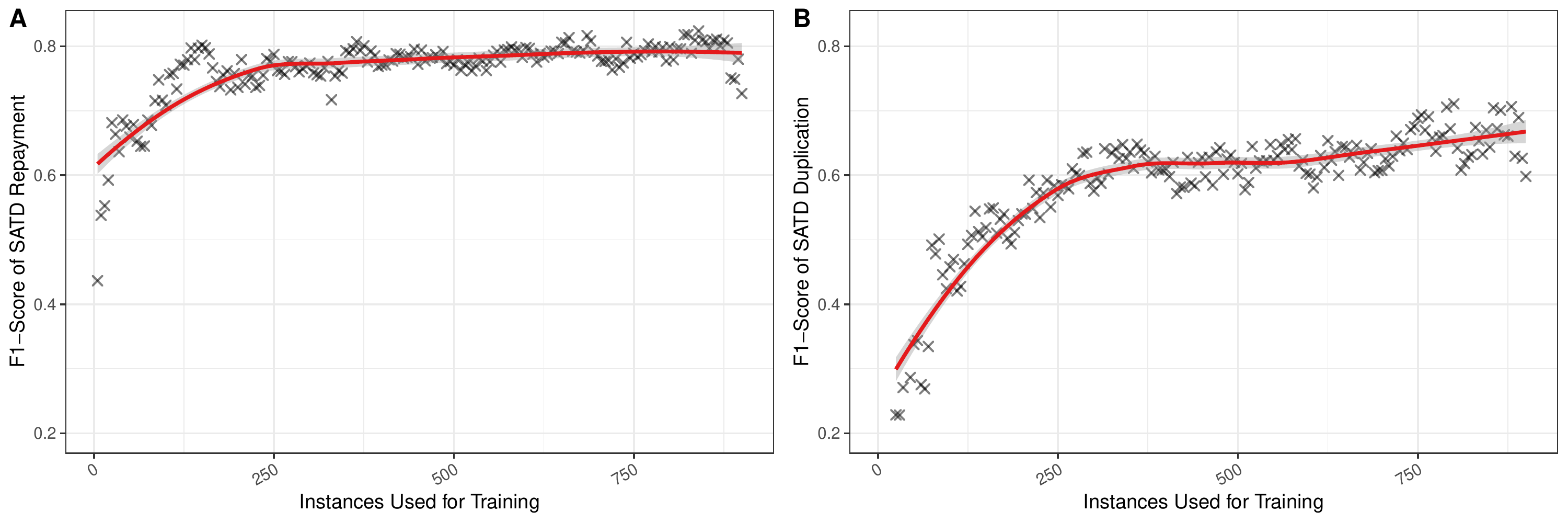}
  \caption{F1-score achieved by increasing the size of the training dataset.}
  \label{f:num_instance_training}
  \vspace{2mm}
\end{figure*}

\begin{framed}
\noindent \textit{The BERT-based approach achieves the highest F1-score of 0.759 and 0.853 for identifying SATD duplication and SATD repayment respectively, with an average F1-score of \textbf{0.806}.}
\end{framed}

\subsubsection{\textbf{RQ1.2: How to further improve the predictive performance of the machine learning model?}}
\label{sec:1.2}

While the BERT-based model achieved an average F1-score of 0.806, there are still opportunities for improvement in identifying duplicated SATD, as evidenced by the lower F1-score of 0.759.
One way is to enhance the model pre-training.

The BERT model is pre-trained on a general-domain corpus, specifically a large corpus of text data comprised of the BookCorpus and English Wikipedia.
However, in-domain pre-training can significantly enhance the predictive performance of the BERT model for text classification \cite{sun2019fine}.
Thus, we explore the effectiveness of pre-training the BERT model on the SATD data.
Because this task focuses on identifying SATD relations and we need domain-specific data to pre-train the BERT model, we first need to collect a sufficient amount of SATD comments for pre-training the BERT model.
Specifically, we utilize the pre-trained machine learning model from the previous work of Li \textit{et al.} \cite{li2022automatic} to collect SATD items from code comments, pull requests, issues, and code commits; these are the raw data without labels.
We then further pre-train BERT with masked language models on the collected SATD comments.
The collected SATD comments for pre-training are made publicly available\footnotemark[1].

\begin{table}[htb]
\caption{Comparison of the predictive performance of BERT models pre-trained on SATD texts with different learning rates (LR).}
\label{tb:improved}
\begin{center}
\resizebox{\columnwidth}{!}{
\def\arraystretch{1.35}
\begin{tabular}{C{1.65cm}C{2.2cm}C{0.7cm}C{0.7cm}C{1cm}}
\hline
\textbf{Pre-Train} & Type of Relation & Pre. & Rec. & F1-score \\
\hline
\multirow{3}{*}{\makecell[c]{\textbf{w/o pre-train} \\(BERT\textsubscript{BASE})}} & SATD Duplication & 0.786 & 0.743 & 0.759 \\
& SATD Repayment & 0.839 & 0.870 & 0.853 \\
\cline{2-5}
& Average & 0.813 & 0.807 & 0.806 \\
\hline
\multirow{3}{*}{\makecell[c]{\textbf{lr: 2e-5} \\(BERT\textsubscript{BASE})}} & SATD Duplication & 0.788 & 0.797 & 0.790 \\
& SATD Repayment & 0.876 & 0.862 & 0.868 \\
\cline{2-5}
& Average & \textbf{0.832} & \textbf{0.830} & \textbf{0.829} \\
\hline
\multirow{3}{*}{\makecell[c]{\textbf{lr: 5e-5} \\(BERT\textsubscript{BASE})}} & SATD Duplication & 0.798 & 0.785 & 0.785 \\
& SATD Repayment & 0.850 & 0.872 & 0.859 \\
\cline{2-5}
& Average & 0.824 & 0.829 & 0.822 \\
\hline
\multirow{3}{*}{\makecell[c]{\textbf{lr: 1e-4} \\(BERT\textsubscript{BASE})}} & SATD Duplication & 0.757 & 0.724 & 0.733 \\
& SATD Repayment & 0.775 & 0.835 & 0.801 \\
\cline{2-5}
& Average & \underline{0.766} & \underline{0.780} & \underline{0.767} \\
\hline
\end{tabular}
}
\end{center}
\end{table}

A common issue in transfer learning is \textit{catastrophic forgetting} \cite{mccloskey1989catastrophic}: during the process of acquiring new knowledge, previously obtained knowledge is erased.
To avoid this, we follow the suggestion of previous work \cite{sun2019fine} to use lower learning rates (i.e., \emph{2e-5}, \emph{5e-5}, and \emph{1e-4}).
The results of different pre-trained models are shown in \cref{tb:improved}; note that reported results are averaged over three random initializations.
We find that two out of three pre-trained models outperform the original BERT-base model (see row \emph{w/o pre-train} in \cref{tb:improved}) in both identifying SATD duplication and SATD repayment.
Generally, using a lower learning rate for pre-training models leads to higher predictive performance.
Specifically, we observe that the pre-trained model with a learning rate of \emph{2e-5} obtains the highest average F1-score of 0.829, while the model with a learning rate of \emph{1e-4} achieves the lowest average F1-score of 0.767.

\begin{framed}
\noindent \textit{After pre-training the BERT-based model, the average F1-score for identifying SATD duplication and SATD repayment is improved from \textbf{0.806} to \textbf{0.829}.}
\end{framed}

\subsubsection{\textbf{RQ1.3: How much data is sufficient for training the machine learning models to identify related SATD?}}
\label{sec:1.3}

To answer this research question, we train the BERT-based model on datasets of various sizes using the settings of the BERT-based model from \cref{sec:1.1}.
Specifically, we first shuffle and split the manually-analyzed dataset into two partitions in a ratio of 1:9.
Then, we use the large partition for training and the small partition for testing.
Because we want to train the model on datasets of various sizes, we create an empty training dataset, add five items to it each time, and train the machine learning model using the training data.

The results of the F1-score for identifying SATD repayment and SATD duplication, achieved while increasing the size of the training dataset, are presented in \cref{f:num_instance_training}; the left subfigure (A) concerns SATD repayment while the right subfigure (B) concerns duplication.
In subfigure (A), we can observe that the BERT-based model can obtain a relatively good F1-score when trained on a very small training dataset.
When more data is added to the training dataset, the model keeps improving until there are around 200 instances.
After that, the model improves rather slowly when further increasing the training dataset, until the curve flattens completely.
In subfigure (B), we find that, the BERT-based model performs much worse when trained on a very small training dataset, compared to identifying SATD repayment relations.
When training on the dataset from 0 to 300 instances, the F1-score goes up dramatically.
After 300 instances, the F1-score improves slowly until there are around 800 instances.
There is a very slight upward trend for the full partition of the training dataset (900 instances).

\begin{framed}
\noindent \textit{To reach a high F1-score, it is sufficient to train the model with approximately 300 instances for identifying SATD duplication and 200 instances for identifying SATD repayment.
The current dataset seems to achieve the highest F1-score for SATD repayment, while for SATD duplication, further increasing the dataset could potentially improve the F1-score, but only slightly. 
}
\end{framed}

\subsection{RQ2: What are the sequences and quantities of SATD relations?}

\subsubsection{\textbf{RQ2.1: What are the sequences of documenting related SATD in different sources?}}
\label{sec:2.1}

To investigate the sequences in which related SATD items are documented, we grouped our SATD relation dataset by source and eliminated cases where there were fewer than five related SATD items.
Through this analysis, we observed nine major cases where the related SATD is documented in other sources. Note that `R' and `D' denote `Repayment' and `Duplication' respectively.
\begin{enumerate}
    \item \textbf{Issue:summary/Pull:summary \textrightarrow~Commit (R)}
    
    Developers create issues or pull requests to report SATD items.
    When this kind of SATD is solved, they always document the repayment of SATD in the commit messages (in contrast this does not always happen when SATD is in the issue comments - see following case).
    For example, a developer wanted to improve the performance of an iterator, so they created a new issue:
    
    \begin{displayquote}
    \textit{``Improve MergeIterator performance''} - [Cassandra-issue-summary-8915]
    \end{displayquote}

    When the improvement is accomplished, it is then documented in the commit message:
    
    \begin{displayquote}
    \textit{``Introduce a more efficient MergeIterator.''} - [Cassandra-commit]
    \end{displayquote}

    Similarly, developers also document SATD in pull requests and then log the repayment in commit messages.
    For example, a developer reported a problem with unnecessary calls in a new pull request:
    
    \begin{displayquote}
    \textit{``Avoid unnecessary leaderFor calls when ProducerBatch queue empty.''} - [Kafka-pull-summary-7196]
    \end{displayquote}

    After solving this SATD, they mentioned the repayment in the relevant commit message:
    
    \begin{displayquote}
    \textit{``Avoid unnecessary leaderFor calls when ProducerBatch queue empty (\#7196).''} - [Kafka-commit]
    \end{displayquote}

    \item \textbf{Issue:comment/Pull:comment \textrightarrow~Commit (R)}

    When SATD is documented in issue comments or pull request comments during the discussion, developers sometimes record its repayment in commits.
    For instance, a developer pointed out a piece of code that is not being used:
    
    \begin{displayquote}
    \textit{``The `unused-argument' is quite popular in tests...''} - [Airflow-pull-comment-5586]
    \end{displayquote}

    When the unused code is cleaned up, it is then recorded in the commit message:
    
    \begin{displayquote}
    \textit{``Remove unused arguments in tests''} - [Airflow-commit]
    \end{displayquote}
    
    \item \textbf{Issue:summary/Pull:summary \textrightarrow~Comment (D/R)}

    Developers occasionally introduce technical debt in new issues or pull requests during software development.
    Subsequently, the same SATD item may be duplicated in other sources, such as code comments.
    For example, a temporary solution was adopted when solving the following Helix issue:
    
    \begin{displayquote}
    \textit{``Enable periodic rebalance as a temporary work-around for the Helix issue.''} - [Pinot-pull-summary-6989]
    \end{displayquote}

    After introducing this technical debt, developers documented this debt as a \emph{TODO} comment in the source code:
    
    \begin{displayquote}
    \textit{``TODO: Enable periodic rebalance per 10 seconds as a temporary work-around for the Helix issue.''} - [Pinot-code-comment]
    \end{displayquote}

    Documenting the repayment of issue SATD in code comments is also common.
    For instance, a developer pointed out a consistency problem in a new issue: 
    
    \begin{displayquote}
    \textit{``Maintained consistency between local, remote, and caching table.''} - [Samza-pull-summary-555]
    \end{displayquote}

    When the debt was solved, they documented the repayment of the debt in a code comment:
    
    \begin{displayquote}
    \textit{``Maintains naming consistency''} - [Samza-code-comment]
    \end{displayquote}
    
    \item \textbf{Issue:comment/Pull:comment \textrightarrow~Comment (D/R)}

    When SATD is discussed in issue comments and developers postpone revolving them in a future iteration, they may document the same SATD in code comments.
    For example, a developer decided to keep using the old variable names but noted that this could be improved in the future, so he created an issue comment:
    
    \begin{displayquote}
    \textit{``I added this TODO. Kept the old config names from BoundedBBPool for BC.''} - [HBase-issue-comment-15525]
    \end{displayquote}

    Then, a relevant TODO comment was added to the codebase.
    
    \begin{displayquote}
    \textit{``// TODO better config names?''} - [HBase-code-comment]
    \end{displayquote}

    In other cases, when SATD is pointed out by developers, they sometimes solve it directly and document the repayment of it in code comments.
    For example, a developer indicated that Java's built-in steam is much slower than loops in an issue comment:
    
    \begin{displayquote}
    \textit{``Nit: streams are significantly slower than for loops even with JDK 12.''} - [Beam-pull-comment-9374]
    \end{displayquote}

    Following that, the debt was solved and documented in a code comment.
    
    \begin{displayquote}
    \textit{``Use a loop here due to the horrible performance of Java Streams''} - [Beam-code-comment]
    \end{displayquote}
    
    \item \textbf{Comment \textrightarrow~Issue:summary/Pull:summary (D)}

    When developers decide to fix SATD which is documented in code comments, they could create an issue or a pull request for it.
    For example, a code comment indicated that this code should be removed after the server is updated: 
    
    \begin{displayquote}
    \textit{``// TODO: remove the code for backward compatible after server updated to the latest code.''} - [Pinot-code-comment]
    \end{displayquote}

    Subsequently, a developer duplicated this SATD item by creating a new issue:
    
    \begin{displayquote}
    \textit{``...after server update to the latest version, remove the code for backward compatible to get the highest performance.''} - [Pinot-pull-summary-910]
    \end{displayquote}
    
    \item \textbf{Comment \textrightarrow~Issue:comment/Pull:comment (D)}
    
    After documenting an SATD item in code comments, developers may sometimes duplicate it in issue or pull comments.
    For example, a developer admitted an SATD item in code comments by stating:
    
    \begin{displayquote}
    \textit{``// TODO: Include the generated file name in the response to the server''} - [Pinot-code-comment]
    \end{displayquote}
    
    Then they mentioned the documentation of this SATD in a pull request comment:

    \begin{displayquote}
    \textit{``// Add a TODO to include the generated file name in the response to server''} - [Pinot-pull-comment]
    \end{displayquote}
    
    \item \textbf{Comment \textrightarrow~Commit (D/R)}

    When developers document the existence or repayment of SATD in code comments and push those commits to repositories, they sometimes also mention the SATD in relevant commit messages.
    For instance, developers added a \emph{TODO} comment about consistency checking:
    
    \begin{displayquote}
    \textit{``// TODO: need GraphComputer.Exceptions consistency checks''} - [TinkerPop-code-comment]
    \end{displayquote}

    When pushing the code to the repository, they also mentioned the introduction of \emph{TODO} in the commit message.
    
    \begin{displayquote}
    \textit{``Add todo for consistency checks for GraphComputer exceptions.''} - [TinkerPop-commit]
    \end{displayquote}

    In addition, the repayment of SATD documented in code comments is also noted in commit messages.
    For example, a developer created a code comment to indicate unnecessary caching that was optimized:
    
    \begin{displayquote}
    \textit{``// avoid unnecessary caching of input...''} - [SystemDS-code-comment]
    \end{displayquote}

    The repayment of this SATD was also recorded in a commit message:
    
    \begin{displayquote}
    \textit{``Improved rdd checkpoint injection (avoid unnecessary input caching)''} - [SystemDS-commit]
    \end{displayquote}
    
    \item \textbf{Comment[Deleted] \textrightarrow~Commit (R)}

    When SATD is repaid and its corresponding code comments are subsequently removed, developers occasionally document the repayment of SATD in commit messages.
    For example, one code comment mentioned the temporary workaround used to initialize static variables:

    \begin{displayquote}
    \textit{``[Deleted] // Servlet injection does not always work for servlet container. We use a hacking here to initialize static variables at Spring wiring time.''} - [CloudStack-code-comment]
    \end{displayquote}

    After this SATD was repaid, the developer documented the repayment of SATD in a commit message:
    
    \begin{displayquote}
    \textit{``Remove temporary hacking and use Official way to wire-up servlet with injection under Spring.''} - [CloudStack-commit]
    \end{displayquote}
    
    \item \textbf{Issue:summary \textrightarrow~Pull:summary (D)}

    After an issue is created to solve SATD items, it is common to create a pull request with a similar description to discuss the detailed solution for fixing the SATD.
    For instance, a developer created an issue to report the problem about expensive tests:
    
    \begin{displayquote}
    \textit{``Speed Up Unit Tests''} - [Drill-issue-summary-5752]
    \end{displayquote}

    Then, they created another pull request to submit solutions and review code to solve this SATD:
    
    \begin{displayquote}
    \textit{``Speed Up Unit Tests add Test Categories''} - [Drill-pull-summary-940]
    \end{displayquote}
\end{enumerate}

\begin{framed}
\noindent \textit{There are \textbf{nine} major cases in which related SATD is documented in a second source, either to duplicate it or report its repayment.
}
\end{framed}

\subsubsection{\textbf{RQ2.2: How much SATD is related between different sources?}}
\label{sec:2.2}

To answer this research question, we first train our BERT-based machine learning model using the best settings described in \cref{sec:1.2}.
We then use the 103 Apache projects described in \cref{sec:DataCollection} to identify the related SATD item pairs using the trained machine learning model.

\begin{table}[htb]
\caption{Number and percentage of SATD items related to different sources.}
\label{tb:num_satd}
\begin{center}
\resizebox{\columnwidth}{!}{
\def\arraystretch{1.35}
\begin{tabular}{p{0.05cm}p{4.28cm}C{0.78cm}C{0.5cm}C{0.78cm}C{0.5cm}}
\hline
\multirow{2}{*}{\hspace{-2mm}\makecell{Major\\Case}} & \multirow{2}{*}{\hspace{1mm}Original \textrightarrow~Duplicated / Repaid} & \multicolumn{2}{c}{Duplication} & \multicolumn{2}{c}{Repayment} \\
\cline{3-6}
 & & \# & \% & \# & \% \\
\hline
3 & \textbf{issue:summary \textrightarrow~comment[added]} & 2,901 & 6.2 & 5,781 & 11.9 \\
1 & \textbf{issue:summary \textrightarrow~commit} & 383 & 0.8 & 1,976 & 4.1 \\
 & issue:summary \textrightarrow~pull:comment & 677 & 1.5 & 518 & 1.1 \\
9 & issue:summary \textrightarrow~pull:summary & 741 & 1,6 & 413 & 0.9 \\
\hline
3 & \textbf{pull:summary \textrightarrow~comment[added]} & 651 & 1.4 & 1,858 & 3.8 \\
1 & \textbf{pull:summary \textrightarrow~commit} & 32 & 0.1 & 1,733 & 3.6 \\
 & pull:summary \textrightarrow~issue:comment & 20 & 0.0 & 208 & 0.4 \\
 & pull:summary \textrightarrow~issue:summary & 43 & 0.1 & 37 & 0.1 \\
\hline
4 & \textbf{issue:comment \textrightarrow~comment[added]} & 25,807 & 55.3 & 15,948 & 33.0 \\
2 & issue:comment \textrightarrow~commit & 64 & 0.1 & 815 & 1.7 \\
 & issue:comment \textrightarrow~pull:comment & 777 & 1.7 & 163 & 0.3 \\
 & issue:comment \textrightarrow~pull:summary & 39 & 0.1 & 90 & 0.2 \\
\hline
4 & \textbf{pull:comment \textrightarrow~comment[added]} & 10,020 & 21.5 & 6,153 & 12.7 \\
2 & pull:comment \textrightarrow~commit & 36 & 0.1 & 966 & 2.0 \\
 & pull:comment \textrightarrow~issue:comment & 62 & 0.1 & 154 & 0.3 \\
 & pull:comment \textrightarrow~issue:summary & 6 & 0.0 & 16 & 0.0 \\
\hline
7 & \textbf{comment[added] \textrightarrow~commit} & 836 & 1.8 & 4,855 & 10.0 \\
6 & \textbf{comment[added] \textrightarrow~issue:comment} & 1,527 & 3.3 & 1,162 & 2.4 \\
5 & comment[added] \textrightarrow~issue:summary & 55 & 0.1 & 200 & 0.4 \\
6 & comment[added] \textrightarrow~pull:comment & 732 & 1.6 & 340 & 0.7 \\
5 & comment[added] \textrightarrow~pull:summary & 57 & 0.1 & 149 & 0.3 \\
\hline
8 & \textbf{comment[deleted] \textrightarrow~commit} & 373 & 0.8 & 3,968 & 8.2 \\
 & comment[deleted] \textrightarrow~issue:comment & 518 & 1.1 & 506 & 1.0 \\
 & comment[deleted] \textrightarrow~issue:summary & 20 & 0.0 & 138 & 0.3 \\
 & comment[deleted] \textrightarrow~pull:comment & 240 & 0.5 & 73 & 0.2 \\
 & comment[deleted] \textrightarrow~pull:summary & 23 & 0.0 & 176 & 0.4 \\
\hline
\end{tabular}
}
\end{center}
\end{table}

\cref{tb:num_satd} presents the number and percentage of related SATD items across the four data sources.
We note that pairs of sources are highlighted in bold when they have more than 1000 related SATD items.
We also note that this automated analysis covers all cases of related SATD items, compared to the nine major cases of the manual analysis discussed in \cref{sec:2.1}; to distinguish these nine major cases, we label them in the first column, according to their numbers in \cref{sec:2.1}.
As we can see, most related SATD items are duplicated/repaid in code comments from various original sources.
Specifically, we observe that SATD items are commonly duplicated/repaid in code comments, starting from issue comments, pull comments, issue summaries, and pull summaries (41755, 16173, 8682, and 2509 respectively).
Additionally, we note that when starting from issue and pull comments, there is a higher likelihood of duplicating SATD in code comments, whereas from issue and pull summaries, there is more SATD repayment in code comments.

After code comments, the commit message is the second most popular source for documenting duplicated/repaid SATD items.
Specifically, a large number of related SATD items are documented in commit messages based on information from added code comments, deleted code comments, issue summaries, and pull summaries (5691, 4341, 2359, and 1865 respectively).
In contrast to documenting related SATD in code comments, the majority of related SATD in commit messages pertains to SATD repayment rather than SATD duplication.

Furthermore, we can see that a certain number of related SATD items are documented from added code comments to issue comments. %
In addition, a small number of related SATD items are documented in other sources, such as comments and summaries of issues and pull requests.

\begin{framed}
\noindent \textit{Developers predominantly document related SATD items in \textbf{code comments}, \textbf{commit messages}, and \textbf{issue comments}.
Additionally, a small number of related SATD items are documented in comments and summaries of issues and pull requests.
}
\end{framed}

\section{Discussion}
\label{sec:discussion}

Based on the findings of our study, we propose the following implications for researchers:

\begin{itemize}
    \item This study proposes a BERT-based deep learning approach for the automated identification of relations between SATD items from four different sources.
    This subsequently allows investigation of the impact of relations between instances of SATD.
    One possible direction is to explore whether documenting duplicated SATD leads to a higher likelihood of SATD repayment in the future.
    Additionally, the identified SATD relations can be utilized by researchers to conduct further studies to better understand SATD repayment.
    For instance, one could investigate the extent to which SATD repayment is documented, why developers choose to do so, and what the best practices are for documenting related SATD items.
    Such studies can provide a deeper understanding of the nature of SATD, helping to manage it more efficiently.
    
    \item Our approach focuses on identifying relations between \textit{pairs} of SATD items; the approach cannot be used to identify relations among \textit{multiple} SATD items.
    Therefore, we suggest that researchers investigate the relations among multiple SATD items by enhancing our approach, or by developing new approaches.
    
    \item In order to facilitate further research in this field, we make our SATD relation dataset and trained machine learning models publicly available\footnotemark[1].
    The former includes 1,000 pairs of SATD items, of which 197 and 390 pairs are classified as SATD duplication and SATD repayment, respectively.
    Sharing the dataset with the research community enables other researchers to build on and improve the work that we have done, and to develop new methods and tools for identifying SATD relations.
    
    \item Our study found that relatively small datasets can achieve decent accuracy in identifying relations between SATD items across different sources (see \cref{sec:1.3}).
    However, we also observed that it is relatively more challenging for machine learning models to capture SATD duplication as compared to SATD repayment (see \cref{f:num_instance_training}).
    Therefore, we recommend that researchers investigate approaches to further improve the accuracy of identifying SATD duplication by augmenting the dataset specifically for SATD duplication or by adopting other machine learning techniques.
    
    \item While our study has explored the generalizability of our approach across 103 open-source projects, it is important to note that the scope of our study is still limited.
    Thus, we recommend that researchers investigate the applicability of our approach to other projects, particularly industrial projects. Additionally, if possible, we advise researchers to make their datasets publicly available\footnotemark[1] for use in training new SATD relation identifiers.
\end{itemize}

In addition to the research implications of our study, we also propose implications for software practitioners:

\begin{itemize}
    \item We recommend that tool developers incorporate our SATD relation identifier into their toolkits and dashboards and experiment with it in practice. 
    Specifically, they can use our relation identifier to explore how SATD is propagated and accumulated through different sources, and look for insights for developers and managers.
    
    \item Our study presents the sequences of documenting related SATD in different sources, which can help practitioners to gain a better understanding of how SATD is created, how it spreads across different sources, and how it is being paid back.
    This information can help practitioners develop more effective strategies for managing SATD and make more informed decisions about how to prioritize their efforts in reducing technical debt.
\end{itemize}

\section{Threats to Validity}
\label{sec:validity}

\subsection{Threats to Construct Validity}

Threats to Construct Validity concern the correctness of operational measures for the studied subjects.
We observed that our dataset of SATD relations is imbalanced.
To address this, we chose to use precision, recall, and F1-score as evaluation metrics, rather than accuracy alone. 
This is to ensure that the performance of our classifier is evaluated in a comprehensive manner and to avoid bias.

\subsection{Threats to Reliability}

Threats to Reliability concern potential bias from the researchers in data collection or data analysis.
Relations between SATD pairs were identified manually for training machine learning models.
To mitigate this threat, the first and second authors independently annotated a sample of 100 SATD pairs and reached an agreement on the classification through discussion.
Additionally, the level of agreement as measured by Cohen's Kappa coefficient was $+0.785$, indicating \emph{substantial} inter-rater agreement.
Furthermore, all data used in this study are publicly available in the replication package\footnotemark[1], allowing for replication and verification of the results.

\subsection{Threats to External Validity}

Threats to External Validity concern the generalization of findings.
In this study, we identified and predicted SATD relations from 103 open-source projects.
Thus, while our findings may be generalizable to other open-source projects of similar size and complexity to the studied Apache projects, we cannot claim broader generalizability.

\section{Conclusion}
\label{sec:conclusion}

In this research, we presented and evaluated methods for automatically identifying relations between instances of SATD across four distinct sources: code comments, issue trackers, commit messages and pull requests.
To accomplish this, we first gathered a dataset of 1,000 pairs of SATD items from 103 open-source projects.
We then conducted a manual analysis of these pairs of SATD items to identify the relations between them.
Using this dataset for training, we compared the predictive performance of two deep learning approaches, BERT-based and CNN-based, and one baseline method, for automatically identifying SATD relations.
Finally, we summarized the characteristics of SATD relations with examples and presented the number of various types of SATD relations in 103 open-source projects.

\bibliography{bibliography}
\bibliographystyle{IEEEtran}

\end{document}